\documentclass[fleqn,usenatbib]{mnras}

\usepackage{newtxtext,newtxmath}

\usepackage[T1]{fontenc}

\DeclareRobustCommand{\VAN}[3]{#2}
\let\VANthebibliography\thebibliography
\def\thebibliography{\DeclareRobustCommand{\VAN}[3]{##3}\VANthebibliography}

\usepackage{graphicx}	
\usepackage{amsmath}	

\newcommand{\kepler}{\textit{Kepler}}

\newcommand{\tess}{\textit{TESS}}
\newcommand{\gaia}{\textit{Gaia}}
\newcommand{\hst}{\textit{HST}}

\newcommand{\galex}{\textit{GALEX}}
\newcommand{\swift}{\textit{Swift}}
\newcommand{\wise}{\textit{WISE}}

\newcommand{\halpha}{$\mathrm{H\alpha}$}
\newcommand{\hbeta}{$\mathrm{H\beta}$}
\newcommand{\hgamma}{$\mathrm{H\gamma}$}
\newcommand{\hdelta}{$\mathrm{H\delta}$}
\newcommand{\thalf}{$t_{1/2}$}
\newcommand{\teff}{$\mathrm{T_{eff}}$}

\title[Multi-wavelength Observations of CR Dra]{A Dragon's Flame of Many Colours: Multi-wavelength Observations of Flares from the Active M Binary CR Draconis}

\author[J. A. G. Jackman et al.]{
James A. G. Jackman$^{1}$\thanks{E-mail: jamesjackmanastro@gmail.com (JAGJ)},
Evgenya L. Shkolnik$^{1}$, 
R. O. Parke Loyd$^{2}$,
Tyler Richey-Yowell$^{3}$\thanks{Percival Lowell Postdoctoral Fellow},
Joe Llama$^{3}$, \newauthor
David Boyd$^{4, 5, 6}$,
Bob Buchheim$^{4, 6}$,
David Iadevaia$^{4, 5}$
Jack Martin$^{4, 5}$, 
Forrest Sims$^{4, 5, 6, 7}$,
Gary Walker$^{4, 6}$,\newauthor
John Wetmore$^{4, 6}$
\\
$^{1}$School of Earth and Space Exploration, Arizona State University, Tempe, AZ 85287\\
$^{2}$ Eureka Scientific, 2452 Delmer Street Suite 100, Oakland, CA, 94602-3017, USA \\
$^{3}$ Lowell Observatory, 1400 W Mars Hill Rd, Flagstaff, AZ, 86001, USA\\
$^{4}$ Spectroscopy Discussion Group (SDG)\\
$^{5}$ British Astronomical Association (BAA), 5 Farringdon Street, London, EC4A 4AB, UK\\
$^{6}$ American Association of Variable Star Observers, 49 Bay State Rd, Cambridge, MA 02138, USA\\
$^{7}$ Astronomical Ring for Access to Spectroscopy (ARAS) \\
}

\date{Accepted XXX. Received YYY; in original form ZZZ}

\pubyear{2015}

\begin{document}
\label{firstpage}
\pagerange{\pageref{firstpage}--\pageref{lastpage}}
\maketitle

\begin{abstract}
We present the results of a multi-wavelength \textcolor{black}{Pro-Am} campaign to study the behaviour of flares from the active M1.5V star binary CR Draconis. CR Dra was observed with \tess\ 20-s photometry, \swift\ near-UV (NUV) grism spectroscopy and with ground-based optical photometry and spectroscopy from a global collaboration of amateur astronomers. We detected 14 flares with \tess\ and \swift\ \textcolor{black}{simultaneously}, one of which \textcolor{black}{also} had simultaneous ground-based photometry and spectroscopy. 
We used the simultaneous two-colour optical and NUV observations to characterise the temperature evolution of the flare and test the accuracy of using optical data to predict NUV emission. We measured a peak temperature of $7100^{+150}_{-130}$\,K for this flare, cooler than the typically assumed 9000\,K blackbody model used by flare studies. We also found that the 9000\,K blackbody overestimated the NUV flux for other flares in our sample, which we attributed to our \swift\ observations occurring during flare decays, highlighting the phase-dependence for the accuracy of flare models.
\end{abstract}




\section{Introduction} \label{sec:intro}
The ultraviolet (UV) emission of stellar flares has captured the attention of the exoplanet research community in recent years. Near-ultraviolet (NUV; 2000--3000\AA) flare emission may drive chemical reactions that form \textcolor{black}{Ribonucleic acid} (RNA) precursors on planets around low-mass stars \citep[e.g.][]{Rimmer18}. The same NUV emission may also drive atmospheric hazes, complicating searches for any present biosignatures. Far-ultraviolet (FUV;1150--1700\AA) photons from flares can dissociate biosignatures such as $\mathrm{H_{2}O}$ and HCN in planetary atmospheres \citep[e.g.][]{Rimmer19}. 

Magnetic reconnection events in the outer atmospheres of stars releases energy to their surroundings, accelerating charged particles along the newly reconnected field lines \citep[e.g.][]{Benz10}. 
These relativistic particles deposit their energy in the chromosphere, heating and evaporating the surrounding plasma, and heating regions on the photosphere \citep[e.g.][]{Watanabe17,Kowalski19}. The heating of the lower stellar atmospheric layers results in intense optical and UV emission. Flares have been observed to increase the UV emission from low-mass stars by factors of hundreds to thousands \citep[][]{Loyd18,MacGregor21}, raising questions about the atmospheric evolution and habitability of orbiting rocky planets \citep[e.g.][]{Venot16}. Information about how often flares occur and their UV spectral characteristics is therefore essential to our understanding how planetary atmospheres evolve.

Studies of flare UV emission require spectroscopic observations from space-based telescopes. New observations are often limited to 
individual stars at a time, performing UV spectroscopy with the 
\textcolor{black}{Space Telescope Imaging Spectrograph} (STIS) or \textcolor{black}{Cosmic Origins Spectrograph} (COS) instruments on the \textit{Hubble Space Telescope} (\hst). These observing campaigns have detected strong UV line emission during flares \citep[][]{Hawley07} and FUV pseudo-continuum temperatures of 15,500\,K and even up to 40,000\,K \citep[][]{Loyd18,Froning19}. The pseudo-continuum is the blackbody spectrum that best describes the emission from combined continuum and line emission. In the UV, flares from magnetically active low-mass stars can change their brightness by several magnitudes in minutes \citep[e.g.][]{Loyd18}. As a result of this, 
these stars must pass brightness checks in both the NUV and FUV in order to preserve \hst\ observing instruments. This limits the number of stars that can be observed with \hst. Active stars that pass these requirements are fainter than some observed in the past, such as AD Leo \citep[e.g.][]{Hawley03}. These restrictions of active M stars 
limit multi-wavelength flare observing campaigns, where flare detection needs to be achieved in multiple bandpasses to build a panchromatic understanding of flare emission \citep[e.g.][]{Tristan23}. While a flare can change the brightness of a low-mass star by magnitudes in the UV, the contrast between the flare and quiescent star is reduced at optical wavelengths. \citet{MacGregor21} used \hst\ STIS to detect a flare that increased the FUV brightness of the M5V star Proxima Centauri by a factor of 14,000. The same flare was found to have a 0.9 per cent amplitude in simultaneous \textcolor{black}{\textit{Transiting Exoplanet Survey Satellite}} (\tess) observations. Low amplitude optical flares are harder to detect on fainter stars, thus further limiting multi-wavelength flare observations with \hst\ and other telescopes. 

In recent years studies have explored the use of the \textit{Neil Gehrels }\textit{Swift} telescope as an alternative to obtaining UV flare observations \citep[][]{Paudel21,Chavali22}. \swift\ is designed for the rapid follow up of gamma ray burst and is equipped with gamma ray, X-ray and UV/optical telescopes. The Ultra-Violet Optical Telescope (UVOT) on board \swift\ can obtain UV or optical observations of designated targets. UVOT is equipped with six filters and two grisms, meaning it can be used for both photometry and spectroscopy.  
Importantly, \swift\ does not have the UV brightness limits imposed on \hst\ observations. Several studies have already combined \swift\ UV photometry with optical photometry (e.g. from \tess) to study relative energies and the morphology of flares in either wavelength regime \citep[e.g.][]{Osten16,Paudel21}. 
\citet{Paudel21} measured the optical and NUV flare rates of EV Lac with \tess\ and \swift, finding that the NUV slope was shallower than for optical flares.

UV photometry has been used to test flare models \citep[e.g.][]{Brasseur19,Jackman22,Brasseur23}, finding that models 
used to describe the flare continuum emission and calculate bolometric energies of events detected with optical observations can underestimate flare UV \textcolor{black}{energies} by up an order of magnitude or more \citep[][]{Jackman22,Brasseur23}. These studies often assume the flare spectrum can be modelled as a single-temperature blackbody, typically with a temperature of 9000-10,000\,K \citep[e.g.][]{Shibayama13,Gunther20,Jackman21}, or a combination of a blackbody spectrum with archival multi-wavelength flare data \citep[e.g. the 1985 AD Leo ``Great Flare''][]{Hawley91, Segura10, Rimmer18}. 
However, photometry is unable to separate line and continuum emission, limiting model tests. This is particularly important in the NUV, where emission lines have been observed to contribute up to 50 per cent of the flare energy budget \citep[][]{Hawley07}. In addition to resolving specific emission lines, NUV spectroscopy can inform us of the shape and strength of the Balmer continuum. Hydrogen recombination radiation has been observed to increase the continuum flux in the NUV to at least twice that predicted by optically calibrated continuum models \citep[][]{Kowalski13,Kowalski19}. The relative flux contribution of this feature also changes with flare phase, increasing in the flare decay. Yet, this feature is largely ignored in the models used to calculate optical and UV flare energies \citep[e.g.][]{Shibayama13} and study the impact of flares on exoplanet habitability \citep[e.g.][]{Rimmer18}. 

The \swift\ UV grism provides a way of circumventing this problem, by offering NUV spectroscopy of targets that may be bright and active. The \swift\ UV grism has a spectral resolution of R$\approx$150 and wavelength coverage of 1700--5000\AA, although order overlap can occur in some spectra at wavelengths above 2740\AA\ \citep[][]{Kuin15}. \citet{Wargelin17} used the \swift\ UV grism to measure the NUV spectra of flares from Proxima Centauri. However, due to contamination in their spectra from nearby stars (the grism dispersion is slitless), they did not perform a full flux calibration of their spectra. More recently, \citet{Chavali22} obtained NUV spectroscopy of a flare from the young active M star AU Mic using the \swift\ UV grism. 
They identified the Mg II h\&k and Fe II emission lines, and constrained the flare temperature to below 15,000\,K. This study highlighted how \swift\ can be used to target  stars that are too bright and active to pass bright object protection checks for \hst.

Bright and active low-mass stars are also ideal targets for ground-based multi-wavelength flare observations, from both professional and amateur observatories. For a long time, amateur observers have been providing high quality photometric and spectroscopic observations to a variety of astronomical studies. These include studies of cataclysmic variables \citep[e.g.][]{Price09}, eclipsing systems \citep[e.g.][]{Osborn19} and more recently stellar flares \citep[][]{Boyd23}. These observations will often also cover longer timespans than those available from professional facilities. This increases the chance of detecting randomly occurring events such as stellar flares in multiple wavelengths, aiding characterisation of their Spectral Energy Distributions (SEDs).

We present the results of a multi-wavelength observing campaign to detect and characterise flares on the active M binary CR Draconis (CR Dra). A spectroscopic binary, CR Dra consists of two active low-mass stars located at a distance of 20.14 pc \citep[][]{BailerJones21,Gaia22}. \citet{Tamazian08} measured an orbital period of 4.04 years using speckle imaging, however \citet{Shkolnik10} and \citet{Sperauskas19} measured periods of 530 days and 1.57 years with radial-velocity observations. 
CR Dra has long been known to undergo flaring activity \citep[][]{Petit57,Cristaldi73}, and it has been studied in multiple wavelengths including radio, optical and X-rays \citep[][]{Tsikoudi97,Callingham21}. CR Dra flares regularly, with \citet{Callingham21} measuring a flare rate of 2.3 flares per day in \tess\ photometry. 
\citet{Welsh06} presented the detection of a flare in 1650s of \galex\ NUV (1771--2836\AA) photometry, within which they found evidence of periodic pulsations. \citet{Million16} later used the gPhoton python package to search for flares in the complete set of archival \galex\ NUV data. They identified eight flares in 49 ks (13.6 hours) of \galex\ data, further showing its active nature. This and its brightness (V=9.5) relative to many other flare stars made it an promising target for characterising the optical and UV properties of stellar flares from low-mass stars.

We obtained simultaneous optical and NUV observations of CR Dra from \tess, \swift\ and a global collaboration of ground-based observatories. We present the instruments used in our observing campaign and then detail how observations were taken. We provide information on how we detected flares in each data set and characterised flares detected in multiple wavelength regimes. Finally, we discuss the results of our campaign and how they aid our understanding of flare emission.

\section{Methods}
We observed CR Dra between the dates of \textcolor{black}{2022-Mar-26} and \textcolor{black}{2022-Jun-21} for a multi-wavelength campaign to study the optical and NUV emission of flares. This campaign combined observations from \tess, \swift\ and ground-based photometry and spectroscopy. A list of start and end dates, along with total exposure times, is given in Tab.\,\ref{tab:exposure_times}.

\begin{table}
    \centering
    \begin{tabular}{c|c|c|c}
         \multicolumn{1}{|p{2cm}|}{\centering Telescope \& \\ Observing Mode} & \multicolumn{1}{|p{1.5cm}|}{\centering Start Date \\ (yyyy-mm-dd)} & \multicolumn{1}{|p{1.5cm}|}{\centering End Date \\ (yyyy-mm-dd)} & \multicolumn{1}{|p{2cm}|}{\centering Total Exposure \\ Time (hrs)} \\
         \hline
         \swift\ UVOT Grism & 2022-04-01 & 2022-06-21 & 20.67\\
         \tess\ & 2022-03-26 & 2022-06-12 & 1695.05* \\
         B-band photometry & 2022-04-14 & 2022-06-21 & 112.47 \\
         Optical spectroscopy & 2022-04-14 & 2022-06-13 & 132.77 \\
    \end{tabular}
    \caption{Start and end dates of the observing campaigns used in this work. The asterisk indicates that we did not use the full \tess\ dataset in our analysis, instead focusing on times with observations by multiple instruments.}
    \label{tab:exposure_times}
\end{table}

\subsection{\tess}
\tess\ observed CR Dra during Sectors 50 to 52 in Cycle 4 as part of Guest Investigation programme G04139 (PI Jackman). These observations took place between 2022-Mar-26 and 2022-Jun-13 and data collection lasted for a total of \textcolor{black}{70.6} days. We have used the publicly available Simple Aperture Photometry (SAP) lightcurves in this work. We used these lightcurves instead of the Pre-search Data Conditioning SAP (PDCSAP) lightcurves in order to maximise the overlap between \tess\ and \swift\ observations. The PDCSAP lightcurves have long term drifts and systematic trends removed, however this can result in data being masked from the final lightcurve. We found that the PDCSAP lightcurve of CR Dra had \textcolor{black}{22} per cent of data points masked. 
To avoid this and maximise the number of simultaneous \tess\ and \swift\ observations we used the SAP flux. 

\subsection{\swift\ UVOT} \label{sec:method_swift_uvot}
Observations of CR Dra were taken with Swift as part of Cycle 18 programme ID 1821143 (PI Jackman). We observed CR Dra with the \swift\ UVOT UV grism in the 0x0384 mode. This mode takes 120s exposures with the UV grism and flushes the buffer with UVM2 observations. \swift\ UV grism observations took place between \textcolor{black}{2022-Apr-01 and 2022-Jun-21}. 

To reduce the UV grism data we followed the methods outlined in \citet{Kuin15} and \citet{Chavali22}, and used the UVOTPY Python package 
\citep[][]{Kuin14}. We used UVOTPY to fit for and extract the first order in each spectrum. During the analysis, UVOTPY fits an anchor position at 2600\AA\ for use in the wavelength calibration. Previous works have noted systematic offsets in wavelength for the fitted anchor position \citep[e.g.][]{Page15}. To account for this we followed \citet{Chavali22} and fitted a Gaussian to the blended Mg II h\&k lines. We shifted each spectrum so that the Gaussian centroid was positioned at 2800\AA. 
As \swift\ uses a slitless grism, 
light from nearby stars can contaminate measured spectra. To account for this we again followed \citet{Chavali22} and visually inspected every 2D grism spectrum. Regions with contamination from light from background stars or order overlap were masked and interpolated over in the final spectra.

We used our grism spectra to generate sets of lightcurves that could be used for flare detection and energy calculation. We first calculated a ``white-light'' lightcurve that integrated between 2200 and 3000\AA. The \swift\ grism has been noted to suffer from order overlap at wavelengths above 2740\AA \citep[][]{Kuin15}. As this falls short of the Mg II h\&k lines at 2800\AA, a region important to our analysis, we inspected all UVOT grism data to identify spectra with order overlap. We found that order overlap was minimal between 2740 and 3000\AA\ for our observations. Some of our spectra had contamination from background sources at wavelengths below 2200\AA. Thus, the low signal to noise at wavelengths shorter than 2200\AA\, and the desire to have a uniform data set for our analysis, prompted us to choose 2200\AA\ as our lower wavelength limit. 
We used the 2200--3000\AA\ lightcurve for energy calculation and model testing. We also generated a 2770--2830\AA\ lightcurve for calculating the flare energy emitted by the Mg II h\&k line. Finally we generated a pseudo-continuum lightcurve, designed to isolate the emission from the NUV continuum. This lightcurve is the sum of flux integrated between 2200--2300, 2640--2730 and 2850--3000\AA. We use the term pseudo-continuum in this work as while these regions were designed to isolate the continuum by avoiding contribution from strong Mg II and Fe lines, it still includes flux from weaker lines in these and other species \citep[][]{Dere97}. Despite this, it still provides a useful probe of the relative behaviours of continuum and line emission.

\subsection{Ground-based Observations}
Ground-based observations were taken by a global collaboration of amateur astronomers. 
These data were a collection of optical photometry and spectroscopy. The start and end dates for the full collection of data are provided in Tab.\,\ref{tab:exposure_times}, however we have also provided the start and end times for each observer below. Here we describe each observatory and instrumentation used to provide data in this work. 

\subsubsection{West Challow Observatory} \label{sec:obs_boyd}
B-band Photometry and optical spectroscopy of CR Dra were taken at the West Challow observatory (WCO) between \textcolor{black}{2022-Apr-15} and \textcolor{black}{2022-May-28}. WCO is a private observatory located in Oxfordshire, UK equipped with two 0.3m class Schmidt-Cassegrain Telescopes, one used for BVRI filtered photometry, the other for low resolution (R$\sim$1000) spectroscopy with a Shelyak LISA spectograph and typical wavelength coverage of 3650--7300\AA. Photometric images of CR Dra were bias, dark and flat corrected and instrumental magnitudes obtained by differential aperture photometry relative to an ensemble of nearby comparison stars with known magnitudes. B-band photometry used an Astrodon Johnson-Cousins filter and had a typical cadence of 30 seconds. Spectroscopic images, typically 300 second integrations, were bias, dark and flat corrected, sky background subtracted, and wavelength calibrated using an internal Neon/Argon calibration source. Spectra were corrected for instrumental and atmospheric losses using spectra of a star situated at a similar airmass to the target star with known spectral profile. 
B-band photometry of CR Dra obtained concurrently with spectra enabled the spectra to be calibrated in absolute flux.

\subsubsection{Desert Celestial Observatory}
Photometry and spectroscopy of CR Dra were taken at the Desert Celestial observatory in Gilbert, Arizona, USA on 11 nights between 2022-Apr-13 and 2022-Jun-12. Spectra were acquired using a Shelyak LISA spectrograph (R=1000) and a cooled Sony ICX825 CCD sensor attached to a 0.5m PlaneWave CDK20 on a PlaneWave L500 robotic mount. Spectra had a typical wavelength coverage of 3720--7310\AA. Each observing run began by acquired flats, followed by a Neon/Argon lamp image and then 7 spectra of the Miles \citep[][]{miles11} reference star HD150117. Slewing to the target CR Dra, another Neon/Argon lamp image was acquired.  Then a continuous sequence of 120 second exposures of CR Dra with no delay between exposures other than camera download time until the end of the nights run at which time another Neon/Argon lamp exposure was acquired. The CR Dra sequence was followed by slewing back to the Miles references star for another set of 7 spectra.  All spectroscopy data was reduced using the “Integrated Spectrographic Innovative Software" (ISIS) package.  ISIS uses the flats, a library of bias and darks at the same temperature as the science exposures and a cosmetic defects list to process the Miles reference star.  This raw Miles 1D spectrum is then compared to the Miles catalog spectrum to calculate an atmospheric and instrument response correction.  The CR Dra spectra are processed in the same way plus the newly calculated response correction is applied to the individual 1D target spectra.  Simultaneously, Photometric data was acquired using a piggy-backed 102mm refractor, a Sony ICX694 CCD sensor and Johnson/Cousin V or B filters. 
Johnson Cousin B filter exposures of 60 sec were continuously acquired on 8 nights.  Each exposure was bias, dark and flat-field corrected.  Differential photometry was extracted using AstroImageJ.  Comparison stars were provided by the AAVSO.

\subsubsection{Lost Gold Observatory}
Spectroscopy of CR Dra was taken at the Lost Gold observatory, located in Gold Canyon, Arizona, USA. Observations were taken using a Shelyak ALPY spectrograph, with a spectral resolution of approximately 500, covering a typical wavelength range of 3750--7250\AA. Spectroscopy of CR Dra was taken between the dates of 2022-May-7 and 2022-Jun-13, with a typical observing cadence of 300 seconds. A Neon/Argon Lamp (for wavelength calibration), and a Reference Star spectrum (HD123299, to determine instrument and atmospheric response) were taken at the beginning of each night, again at meridian-flip, and again at the end of a night. Each spectrum was reduced and airmass corrected using the ISIS software package. 

\subsubsection{Sierra Remote Observatory}
B-band photometry of CR Dra was taken at the \textcolor{black}{Sierra Remote} observatory, located in \textcolor{black}{Auberry, California}. Observations were taken using a \textcolor{black}{CDK20 20$''$} telescope with a FLI Kepler-400 camera between the dates of 2022-Apr-14 and 2022-Jun-12. B-band photometry was taken using an Astrodon Johnson-Cousins filter with a typical cadence of 55 seconds. Images were bias, dark and flat-field corrected, and magnitudes obtained through differential aperture photometry with a comparison star of known magnitude. Observations were reduced using the MaxIm DL software package. 

\subsubsection{Huggins Spectroscopic Observatory}
Optical spectroscopy of CR Dra was taken at the Huggins Spectroscopic Observatory in Essex, UK between the dates of \textcolor{black}{2022-Apr-15} and \textcolor{black}{2022-Apr-26}. Spectra were obtained using a Lhires III spectrograph with 1200 l/mm grating (R$\sim$8000) and Atik 460EX CCD camera with a Celestron C14 14'' telescope.
A cadence of 300 seconds was used when acquiring spectra. Spectra were chosen to cover a wavelength range of 4650--5130\AA. This was done to focus on the \hbeta\ emission line, which can exhibit greater flare amplitudes than \halpha\ and thus useful for flare detection \citep[e.g.][]{Muheki20}. Observations were reduced using the ISIS software package.

\subsection{Spectral Energy Distribution Fitting} \label{sec:method_sed_fitting}
To obtain stellar properties we could use in our analysis we fit the Spectral Energy Distribution (SED) of CR Dra. Previous works have assumed equal or similar masses for the two stars, with a shared spectral type of M1.5Ve \citep[][]{Shkolnik10,Callingham21}. However, \citet{Tamazian08} used differential photometry to measure a V magnitude difference of $1.78\pm0.07$ between the two components. This led them to identify spectral types of M0 and M3. This suggests that we should not consider the system as a single source, and that derived flare properties will change depending on the star. To account for this in our analysis, we fit the SED of each star.

We used the PHOENIX v2 stellar models in our fitting \citep[][]{Allard12}. We convolved each PHOENIX model spectrum with photometric bandpasses to create a grid of synthetic photometry in effective temperature space. We modelled the photometry of each star separately and then added them to get the combined fluxes. We then fit the combined synthetic photometry to the measured broadband photometry of CR Dra. We placed a Gaussian prior on the V band magnitude difference in line with the \citet{Tamazian08} result. 
The broadband fluxes used in our fitting were obtained from APASS \citep[][]{Henden14}, 2MASS \citep[][]{Skrutskie06}, \wise \citep[][]{Cutri21}, and \gaia\ DR3 \citep[][]{Gaia22}. Each model was normalised using a scaling factor equal to $(R/d)^{2}$ where $R$ is the stellar radius and $d$ is the distance. We ran our fitting with a Markov Chain Monte Carlo process, using 32 walkers for 50,000 steps. We used the final 10,000 walkers to sample the posterior distribution. We placed a Gaussian prior on the distance, using the values of CR Dra measured by \citet{BailerJones21} from \gaia\ EDR3 astrometry. 

We obtained best fitting effective temperatures of \textcolor{black}{$3550\pm120$}\,K and \textcolor{black}{$3240\pm140$}\,K, corresponding to spectral types of M1.5 and M3 \citep[][]{Cifuentes20}. To account for the two different spectral types, we considered a scenario for each star. We used our best fitting SEDs to calculate the relative flux dilution in the \tess\ and B bandpasses, and used these to adjust the optical flare amplitudes when calculating flare temperatures and testing flare models later in this work.

\subsection{\tess\ Flare Detection and Energy Calculation}

The purpose of this study is to identify flares that had multi-wavelength observations. We therefore focused on identifying flares in the \tess\ lightcurves at times where \tess\ and another observatory were observing CR Dra. We isolated regions in the \tess\ lightcurve up to six hours before and after each \swift\ and ground-based observing block. We visually inspected these \tess\ lightcurves to look for flares. We performed visual inspection over other methods in order to identify low amplitude flare events that may be missed or downweighted by automated methods \citep[e.g.][]{Feinstein20}. 
\citet{Callingham21} 
measured a flare rate of 2.3 flares per day from \tess\ 2-minute cadence photometry. 
We analysed \textcolor{black}{27.57 days} of data out of the total \textcolor{black}{70.6 days} of \tess\ observations. We identified \textcolor{black}{113} flare candidates in this 27.57 days, a rate of 4.1 flares per day from the combined CR Dra system. We attribute the higher flare rate to the use of 20-second over 2-minute cadence photometry. Longer cadence photometry can smear out short duration flares, reducing their observed amplitude and causing some events to be missed entirely by detection algorithms \citep[e.g.][]{Yang18,Howard22}

We visually inspected the flare candidates in the lightcurves and postage stamps to remove false positives due to objects passing through the aperture, or flares from nearby stars \citep[e.g.][]{Jackman21}. We finally visually inspected all the confirmed candidates and manually selected the start and end times of detected flares. Some flares have very impulsive rise and initial decay phases that go above our detection limit, but also feature long gradual decay phases. Using where the flare goes above and below the detection limit as our start and end times would result in us missing these regions when calculating flare energies. We therefore visually inspected each flare and manually set the start and end time.   

Spectroscopy of flares have shown they release energy via line emission and heat rapidly during their impulsive rise phase, and subsequently cool during their decay. Spectroscopy and multi-colour photometry has been used to measure temperatures up to 20,000\,K and even 40,000\,K at flare peaks \citep[][]{Loyd18,Froning19,Howard20}. Despite this, we have elected to use the 9000\,K blackbody to calculate flare energies so that we are able to compare and directly test model assumptions used in the literature \citep[e.g.][]{Jackman22}. We calculated the energy of each flare following the method outlined in \citet{Shibayama13}. We assumed the flare spectrum can be modelled as a 9000\,K blackbody and renormalised the ratio between the flare and stellar spectrum using the measured amplitude in the \tess\ lightcurve. We integrated between the start and end times specified during our visual inspection of flare candidates. The use of a 9000\,K blackbody assumes the heated region in the stellar atmosphere behaves as a perfect emitter with a fixed temperature. 

\subsection{Ground-based Photometry Flare Detection}
We obtained B band photometry from the West Challow, Desert Celestial and \textcolor{black}{Sierra Remote} observatories. 
Data was reduced by each observer and uploaded to the AAVSO International Database\footnote{\url{https://www.aavso.org/webobs/}}. 
These observations were uploaded in UTC time format. To bring them onto the same format as \tess\ observations, we applied a barycentric correction to all observation times using the astropy.time package \citep[][]{astropy:2013,astropy:2018,astropy:2022}. We then converted all times from UTC to Barycentric Dynamical Time (TDB), the format used by \tess. We identified flares in the ground-based photometry through comparison of our data with \tess\ observations taken at the same time. We identified \textcolor{black}{seven} flares that had simultaneous \tess\ and B band photometry. 

\subsection{Ground-based Spectroscopy}
We obtained optical spectroscopy of CR Dra from \textcolor{black}{four} observatories. These were the West Challow, Desert Celestial, Lost Gold and Huggins Spectroscopic observatories. 
We used the spectroscopy from the West Challow, Desert Celestial and Lost Gold observatories to measure equivalent widths and create lightcurves of the \halpha, \hbeta\ and \hgamma\ lines. 
We measured the equivalent width of each line \citep[e.g.][]{Reid95,Newton17} using the integration windows listed in Tab.\,\ref{tab:cont_regions}. These windows were chosen specially in order to encompass the full width of the Balmer lines in our spectra (R=500--1000). Poisson uncertainties were propagated through our analysis and into our final results.

Where available, spectra were flux-calibrated using simultaneous B band photometry. We used these to calculate the average quiescent luminosity of CR Dra in continuum regions around the \halpha, \hbeta, and \hgamma\ lines. We calculated the energy in each emission line by multiplying the equivalent width by the flux-calibrated continuum emission. As spectroscopy was taken simultaneously with \tess\ observations, we used the \tess\ lightcurve to identify quiescent and flare spectra. 

\begin{table}
    \centering
    \begin{tabular}{c|c|c|c}
        Line & Line Window (\AA) & Blue Continuum (\AA) & Red Continuum (\AA) \\
        \hline
        \halpha\ & 6648-6678 & 6500--6545 & 6580--6620 \\
        \hbeta\ & 4853--4870 & 4820--4840 & 4875--4890 \\
        \hgamma\  & 4355--4370 & 4300--4327 & 4330--4350 \\
        \textcolor{black}{\hdelta} & 4094--4110 & 4085--4093 & 4111--4115 \\
        
    \end{tabular}
    \caption{Custom integration windows used for calculating equivalent widths in our spectroscopic analysis.}
    \label{tab:cont_regions}
\end{table}

\subsection{\swift\ UVOT Flare Detection and Energy Calculation} \label{sec:method_swift_energy}
We used the start and end times of flares identified in our \tess\ lightcurve to analyse flare events in our \swift\ observations. 
We calculated the NUV energies of flares in each of the lightcurves discussed in Sect.\,\ref{sec:method_swift_uvot}, to calculate the total NUV energy, energy in the Mg II h\&k line and the energy emitted from the NUV continuum. As our UV spectra are flux-calibrated, the measured energy does not depend on the assumed source. We integrated flux-calibrated lightcurves (with the quiescence subtracted) from Sect.\,\ref{sec:method_swift_uvot} within the \tess\ flare timings and multiplied by 4$\pi d^{2}$, where $d$ is the distance, to measure the UV flare energies.

The \swift\ grism observing strategy results in flares with durations longer than \textcolor{black}{10 minutes} being partially observed. Many flares will only be observed during a single \swift\ visit. However, longer duration events can be covered by consecutive visits. Energies for flares observed in a single visit were calculated by integrating over the times covered by the respective observations.  Flares that were observed over consecutive visits had their energies calculated by integrating over each visit and any time gap in between. We note that as the \swift\ visits do not cover the full duration of flare events, our reported energies are lower limits.

\subsection{Flare Temperatures} \label{sec:method_flare_teff}
Flares with simultaneous \tess\ and B band photometry were used to measure flare temperatures. We fit blackbody curves to the flare flux in each filter to measure pseudo-continuum flare temperatures \citep[e.g.][]{Jackman22}. We use the term pseudo-continuum as the fitted blackbody curves account for  both the continuum and line emission in each filter. As noted by \citet{Jackman22} this fitting may overestimate the flare temperatures, due to the stronger line contribution at blue optical wavelengths. To fit the blackbody temperatures we generated a grid of 1000 blackbody spectra with effective temperatures between 6000 and 50,000\,K. We multiplied each of these by the \tess\ and B bandpass 
to get the flux in each filter \citep[e.g.][]{Howard20}. We also multiplied our best fitting PHOENIX spectrum for each star from Sect.\,\ref{sec:method_sed_fitting} by each filter to calculate the quiescent fluxes. We did this for both stars to account for variations in flare amplitude. 
We divided our grid of blackbody flare fluxes by the quiescent fluxes for each star to generate grids of flare amplitudes. 
We then renormalised these flare amplitudes to have a \tess\ flare amplitude of unity (i.e. the flare and quiescent star emit equal flux in the \tess\ bandpass). 

We followed the methods outlined by \citet{Howard20} to measure the flare temperature associated with different flare phases. We first fit the \citet{Davenport14} flare model to the amplitudes in both the B-band and \tess\ photometry. We did this using an MCMC method with 32 walkers and 20,000 steps, using the final 5000 to sample the posterior probability distribution. We then used the flare model in each bandpass to generate a smoothly varying temperature model.  We performed our fitting with the photometry from the full system, and then adjusted the amplitudes for each star to measure two sets of temperatures. 

We used our models to calculate the global average flare temperature,  temperature during the flare FWHM and the peak flare temperature. 
The first was calculated for comparison with single temperature flare models (e.g. 9000\,K blackbody). The second was to measure the temperature around the flare peak and was used by \citet{Howard20} as a proxy for the peak flare temperature. The third was to sample the maximum temperature. 
We chose a sample of 500 randomly selected models from our posterior distribution and used these to calculate the distribution of temperatures. We then calculated the 16th, 50th and 84th percentiles to estimate the lower and upper uncertainties on each value. Our best fitting temperatures and uncertainties for each star are given in Tab.\,\ref{tab:flare_teff}.

\def\arraystretch{1.2}
\begin{table*}
    \centering
    \begin{tabular}{c|c|c|c|c|}
         Flare Start (TBJD) & Flare End (TBJD) & Global \teff\ (K) & FWHM \teff\ (K) & Peak \teff\ (K) \\
         \hline
2688.856868125173 & 2688.8723774301957 & $8160^{+330}_{-310}$ ($8270^{+450}_{-290}$) & $8890^{+430}_{-330}$ ($8960^{+340}_{-330}$) & $10590^{+1280}_{-890}$ ($10690^{+1210}_{-930}$) \\
2688.920525720016 & 2688.9608036158215 & $7110^{+160}_{-260}$ ($7050^{+280}_{-240}$) & $7050\pm210$ ($7050^{+200}_{-210}$) & $8480^{+550}_{-460}$ ($8480^{+560}_{-480}$) \\
2689.672842648601 & 2689.6823334145715 & $9480^{+1580}_{-1380}$ ($9700^{+1480}_{-1500}$) & $10710^{+1810}_{-1720}$ ($10810^{+1840}_{-1600}$) & $17280^{+8840}_{-3980}$ ($17100^{+9880}_{-4610}$) \\
2689.9779360287534 & 2690.0293250450277 & N/A & $8230^{+35}_{-31}$* ($8250^{+39}_{-31}$*) & $12640^{+110}_{-100}$* ($12680^{+130}_{-110}$*) \\
2707.7679000204157 & 2707.7815572873524 & N/A & $6670^{+500}_{-470}$ ($6650^{+540}_{-470}$) & $9760^{+2410}_{-2060}$ ($9420^{+2370}_{-1850}$) \\
2708.3787735073875 & 2708.456550520614 & $6050^{+70}_{-60}\dagger$ ($6090^{+70}_{-60}$) & $6320^{+50}_{-60}\dagger$ ($6340^{+50}_{-70}$) & $7100^{+150}_{-130}\dagger$ ($7130^{+150}_{-130}$) \\

2708.7799270056357 & 2708.7919639265756 & $7670^{+3920}_{-1670}$ ($7335^{+3930}_{-1335}$) & $7803^{+1570}_{-1090}$ ($7600^{+1640}_{-1130}$) & $10760^{+3730}_{-2110}$ ($10460^{+3710}_{-2240}$) 
    \end{tabular}
    \caption{Temperatures of flares with two-colour photometry in our sample, assuming the flare came from the primary (secondary) star. N/A indicates where a temperature value could not be measured due to a partial observation or low signal to noise in one lightcurve. Asterisks indicate where the ground-based photometry was used over a fitted model in our temperature analysis. The $\dagger$ symbol indicates that this flare was fit using the \citet{Mendoza22} model and is discussed further in Sect.\,\ref{sec:result_may8}.}
    \label{tab:flare_teff}
\end{table*}
\def\arraystretch{1.0}

\section{Results}
We conducted an multi-wavelength observing campaign to study flares from the active M binary CR Dra. We obtained near-ultraviolet spectroscopy with \swift, 20-s cadence red-optical photometry with \tess\ and optical photometry and spectroscopy from ground-based observatories. We observed \textcolor{black}{14} flares with simultaneous \swift\ spectroscopy and \tess\ photometry, and \textcolor{black}{18} flares with \tess\ and ground-based photometry or spectroscopy. Of the 14 flares with simultaneous \swift\ and \tess\ observations, one was caught with ground-based photometry and spectroscopy. 

\subsection{\tess\ and \swift} \label{sec:energy_result}
We identified \textcolor{black}{14} flares in our analysis that had simultaneous \tess\ and \swift\ observations. The \swift\ observations were taken during the flare decay phase for all but \textcolor{black}{one}, which was taken during the flare rise. We used the NUV spectroscopy to measure flare energies for each event. For each NUV spectrum we calculate the energy following the method outlined in Sect.\,\ref{sec:method_swift_energy}. 
We measured 2200--3000\AA\ NUV flare energies between \textcolor{black}{$1.5\pm1.0\times10^{31}$} and \textcolor{black}{$1.2\pm0.2\times10^{33}$} ergs, and Mg II h\&k emission line energies between \textcolor{black}{$2.4\pm1.4\times10^{30}$} and \textcolor{black}{$1.8\pm0.4\times10^{32}$} ergs. We measured that the Mg II h\&k lines emitted between 9 and 24 per cent of the 2200-3000\AA\, energy, with 93 per cent of flares having a fraction less than 20 per cent.

\subsection{Flare Caught Simultaneously With \tess, \swift\ and from the Ground} \label{sec:result_may8} 
\begin{figure}
    \centering
    \includegraphics[width=\columnwidth]{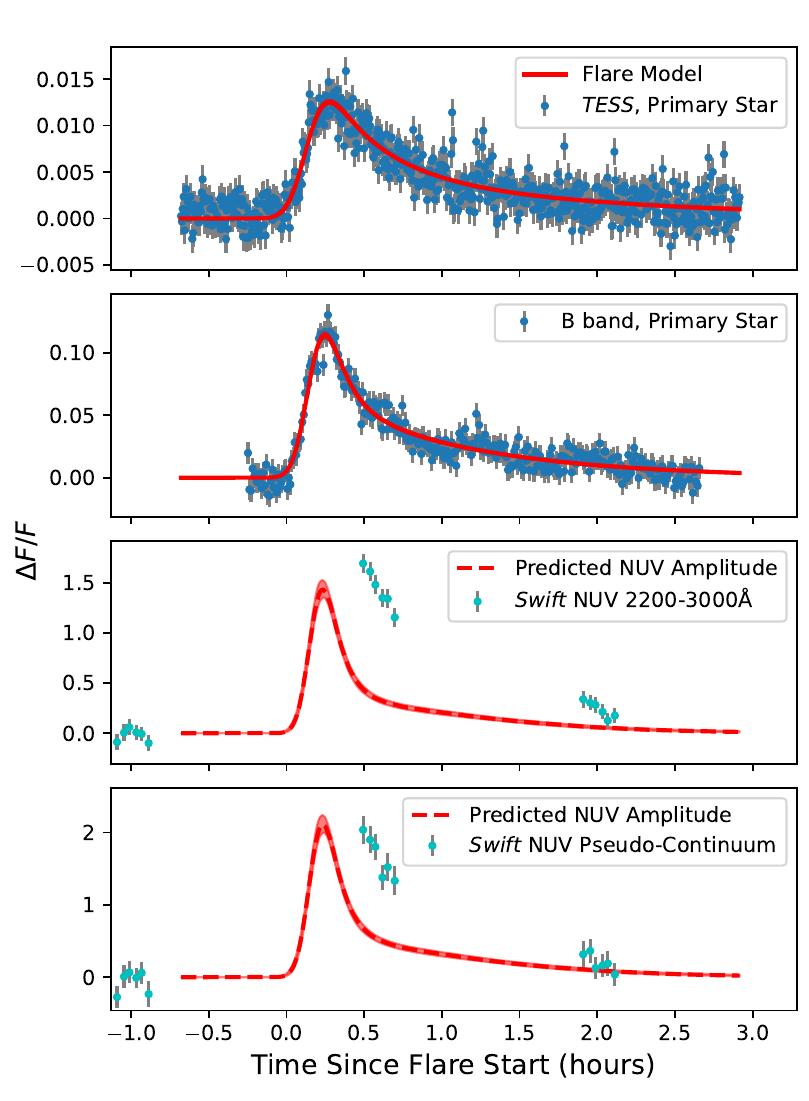}
\caption{\tess, B-band and \swift\ NUV lightcurves of the flare on the 8th May 2022. We fit an empirical flare model to the \tess\ and B-band photometry, which we used to model the temperature variation during the flare. We have assumed here that the flare came from the primary star. This model was used to predict the NUV flux from the optical continuum during the flare (Sect.\,\ref{sec:result_may8}), showing how it underestimates the observed NUV lightcurve. The observed \swift\ NUV amplitude is for the combined system, and is diluted from the true value.} 
    \label{fig:uv_flare}
\end{figure}

We detected one flare with \tess, \swift\ and from the ground on the 8th May 2022. This flare started at \textcolor{black}{21:03:18 UT (MJD 59707.87730)} and lasted for 2.5 hours. This flare was observed with B-band photometry and optical spectroscopy from West Challow Observatory. B-band photometry took place between 20:49:56 UT (MJD 59707.86801) and 23:44:11 UT (MJD 59707.98902), and spectroscopy was taken between 20:58:43 UT (MJD 59707.87412) and 23:33:24 UT (MJD 59707.98154). The B-band photometry was taken with a 35s cadence and the spectroscopy had a 300s cadence. These observations were simultaneous with \tess\ 20-s cadence photometry and \swift\ UV grism spectroscopy. Lightcurves of the flare in the \tess\ bandpass, B-band and the NUV are shown in Fig.\,\ref{fig:uv_flare}. Optical photometry and spectroscopy were recorded during the whole flare, while NUV spectroscopy was obtained during the decay phase. 

We used the B-band photometry to flux calibrate the optical spectra. We did not obtain spectra of the quiescent star on the night of the flare. We 
constructed a quiescent optical spectrum of CR Dra from observations taken on the 27th May 2022. This night was chosen as observations were taken in a similar airmass range as the night of the flare. We normalised this spectrum to the quiescent B band photometry before the flare. The renormalised quiescent spectrum was then subtracted off spectra taken during the flare to obtain flare-only data. 

\begin{figure*} 
    \centering
    \includegraphics[width=\textwidth]{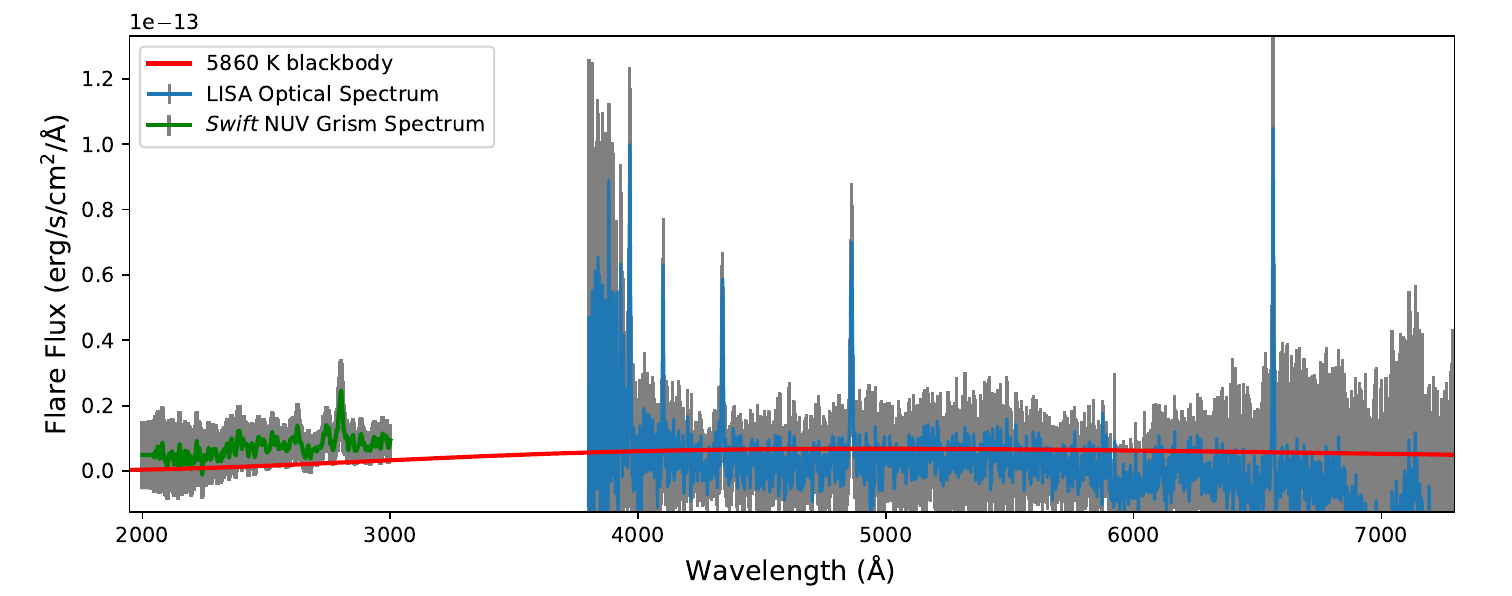}
    \caption{\swift\ NUV and LISA optical spectrum (Sect.\,\ref{sec:obs_boyd}) of the flare emission observed on the 8th May. This is during the first \swift\ visit for this flare. As in Fig.\,\ref{fig:uv_flare} we assume the flare comes from the primary star. The grey regions are the associated 1$\sigma$ uncertainties for each dataset. The red line is a blackbody with temperature from our temperature analysis in Sect.\,\ref{sec:result_may8}, normalised to give the measured \tess\ amplitude during the time of observations. We have masked \swift\ grism data at wavelengths longer than 3000\AA\ due to potential order overlap. We discuss the apparent negative flux at wavelengths greater than 6000\AA\, in Sect.\,\ref{sec:result_may8}.}
    \label{fig:flare_only_spec}
\end{figure*}

\swift\ observed this flare on two consecutive visits. The times of these visits are shown in Fig.\,\ref{fig:uv_flare}. The first occurred near the start of the decay phase and lasted 12 minutes, and the second began \textcolor{black}{72 minutes after this,} at the end of the decay phase. We measured a quiescent NUV spectrum for CR Dra by taking a weighted mean of spectra from the \textcolor{black}{six} visits preceding the flare. We calculated the average spectrum in each visit and subtracted our quiescent spectrum from these to obtain flare-only NUV spectra. We combined these with our simultaneous optical spectra to build two flare-only spectra of this event from CR Dra. The first of these is shown in Fig.\,\ref{fig:flare_only_spec}. We can see in Fig.\,\ref{fig:flare_only_spec} that the first visit shows both continuum and line emission in the NUV, while only line emission appears to be present in the optical spectrum. We note an apparent negative continuum flux at wavelengths longer than 6000\AA. We attribute this to a potential change in starspot coverage between the night of the flare (8th May 2022) and the night used to construct the quiescent spectrum (27th May 2022) \citep[e.g.][]{Rackham18}.

We fit both our \tess\ and B band lightcurves to measure the relative timescales and measure a smoothly varying colour temperature. We fit each lightcurve using the empirical flare template from \citet{Mendoza22}. This template is based on the one from \citet{Gryciuk17} adapted by \citet{Jackman18} and describes the flare as the convolution of a Gaussian heating profile and two exponential decays. We fit each profile using an MCMC process that we ran with 16 walkers for 100,000 steps, taking the final 10,000 to sample the posterior distribution. We used our best fitting models to measure the two-colour lightcurve of the flare. We used this colour to measure the flare temperature following the method outlined in Sect.\,\ref{sec:method_flare_teff}. We measured a peak temperature for the primary (secondary) star of \textcolor{black}{$7100^{+1550}_{-130}$\,K} \textcolor{black}{($7130^{+150}_{-130}$)}, a FWHM temperature of \textcolor{black}{$6320^{+50}_{-60}$}\,K \textcolor{black}{($6340^{+50}_{-70}$)} and a global average temperature of \textcolor{black}{$6050^{+70}_{-60}$\,K} \textcolor{black}{($6090^{+70}_{-60}$)}. The peak temperature of the flare occurred during the rise phase, during the initial energy deposition in the flaring atmosphere. 

We used our two-colour temperature model to calculate the temperature during each \swift\ visit. We measured a two-colour temperature of $5860^{+80}_{-90}$\,K \textcolor{black}{($5880^{+80}_{-90}$)}\,K for the first \swift\ visit, and $5540^{+230}_{-220}$\,K \textcolor{black}{($5580^{+230}_{-220}$)}\,K for the second visit. These values show a slight drop in temperature during the flare decay. 
We then used the two-colour temperature model to predict the NUV emission in our observations. 
We did this for both the 2200--3000\AA\ and the pseudo-continuum NUV emission. The comparison can be seen in Fig.\,\ref{fig:uv_flare}, where we can see the extrapolation of the optical continuum underestimates the observed NUV emission. We calculated that the two-colour temperature model underestimated the 2200-3000\AA\, NUV flux by a factor of \textcolor{black}{$4.2\pm0.2$} \textcolor{black}{($4.1\pm0.2$)} during the first visit and a factor of \textcolor{black}{$4.1\pm0.9$} \textcolor{black}{($4.1\pm0.9$)} during the second. These results highlight the difficulty of predicting UV flare emission from two-colour optical photometry. We also calculated the discrepancy for our pseudo-continuum UV lightcurves, measuring values of \textcolor{black}{$3.1\pm0.2$} \textcolor{black}{($3.1\pm0.2$)} and \textcolor{black}{$2.1\pm1.1$} \textcolor{black}{($2.1\pm1.1$)} for the two visits. The decreased discrepancy for the pseudo-continuum shows the effect of the flux from the Fe II and Mg II lines on measurements of the bulk UV emission. 

\begin{figure} 
    \centering
    \includegraphics[width=\columnwidth]{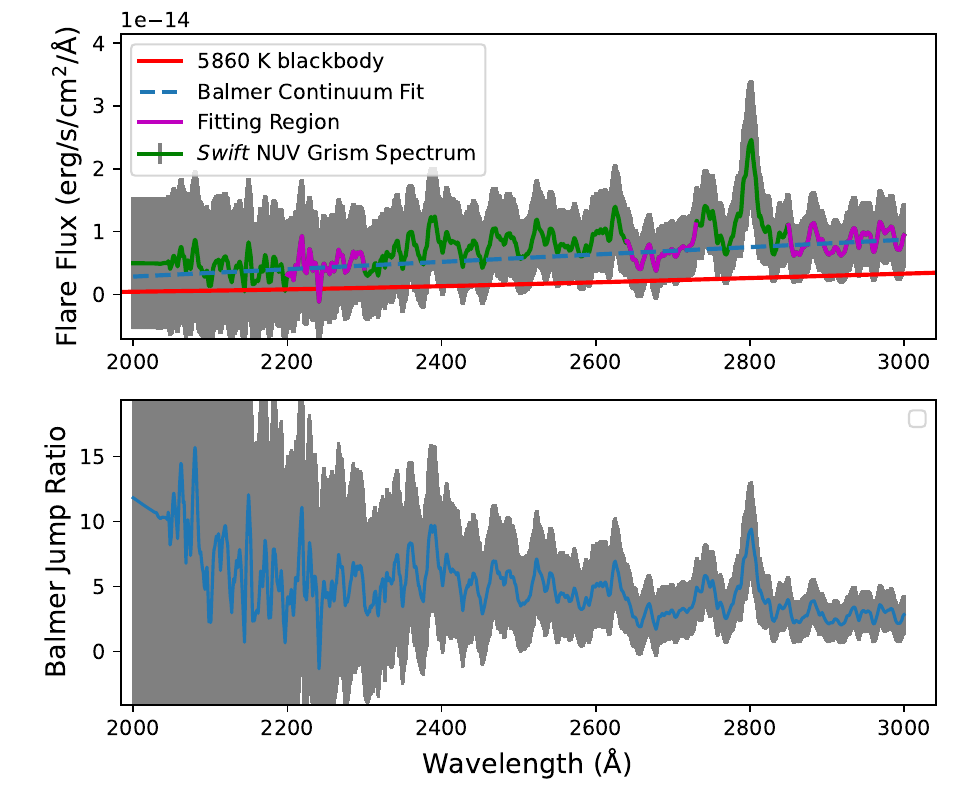}
    \caption{Top: A zoom in of the \swift\ NUV spectrum shown in Fig.\,\ref{fig:flare_only_spec}.
    The grey regions are the associated 1$\sigma$ uncertainties for each dataset. The red line is a blackbody with temperature from our temperature analysis in Sect.\,\ref{sec:result_may8}, normalised to give the measured \tess\ amplitude during the time of observations. As in Fig.\,\ref{fig:uv_flare} we have assumed the flare originated on the primary star. The blue dashed line is a linear fit to the pseudo-continuum regions, shown in purple. Bottom: The Balmer \textcolor{black}{jump ratio}, measured from the ratio between the NUV emission and the best fitting blackbody.}
    \label{fig:balmer_dec}
\end{figure}

Our measurements of the discrepancy in the pseudo-continuum also highlight the contribution from the Balmer continuum. At wavelengths shorter than 3646\AA\, the Balmer series contributes an additional continuum \citep[e.g.][]{Kowalski13}, increasing the NUV emission above that predicted by a blackbody curve fit to the optical spectrum \citep[e.g.][]{Kowalski19}. The ratio between the NUV continuum and the extrapolated optical continuum, the Balmer \textcolor{black}{jump ratio}, has been observed to increase during flare decays. 
We used our NUV spectra and two-colour temperatures with the \tess\ photometry to measure the Balmer \textcolor{black}{jump ratio} in each \swift\ visit. We fit the NUV pseudo-continuum of each grism spectrum with a linear relation and divided this by the corresponding blackbody determined from our two-colour temperature analysis. 
We measured decrements ranging between \textcolor{black}{$2.6\pm0.1$} (\textcolor{black}{$2.6\pm0.1$}) at 3000\AA, and \textcolor{black}{$5.0\pm0.5$} (\textcolor{black}{$4.9\pm0.5$}) at 2200\AA\, for the first visit. These values changed to $1.8\pm0.3$ ($1.7\pm0.3$) at 3000\AA, and $3.2\pm2.9$ ($3.1\pm2.9$) at 2200\AA\, in the second visit. Our fit for the first visit is shown in Fig.\,\ref{fig:balmer_dec}. 
The increasing decrement in both visits is likely driven by the rapidly diminishing flux from the $\approx 5900$\,K blackbody curve as we approach the FUV, rather than an increasing strength of the Balmer jump at shorter wavelengths.

\begin{figure}
    \centering
    \includegraphics[width=\columnwidth]{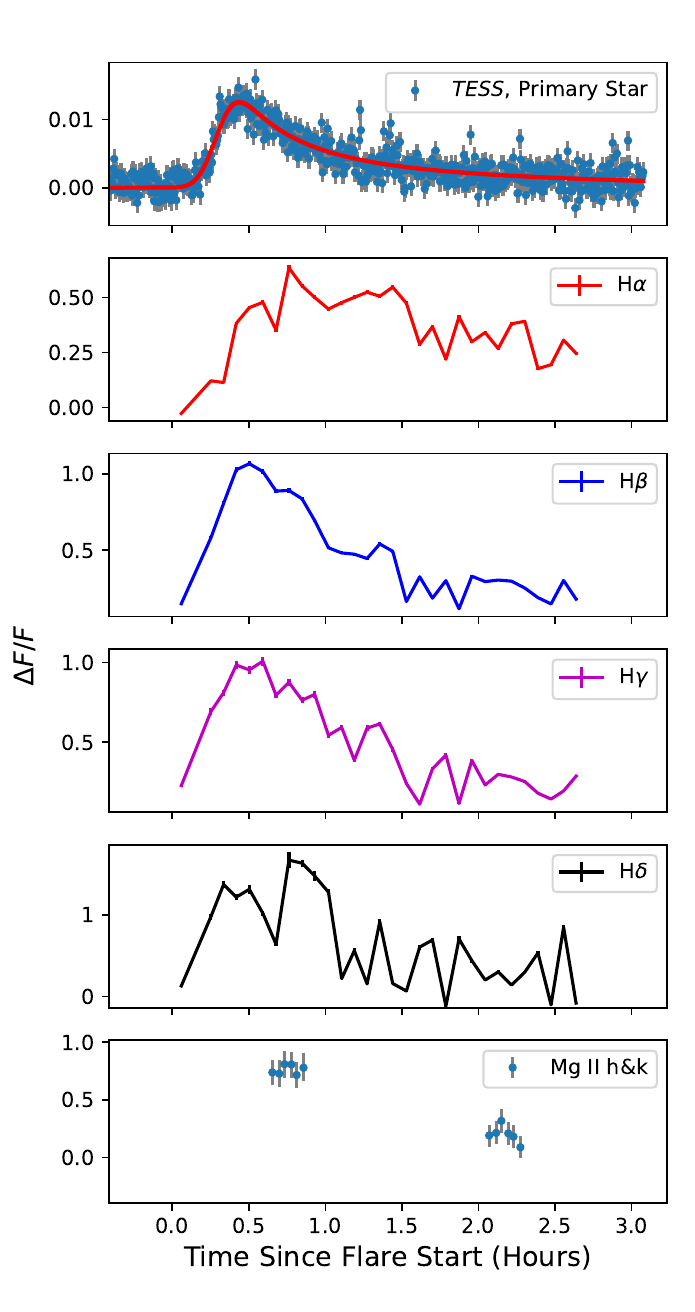}
    \caption{Comparison of the \tess, \halpha, \hbeta, \hgamma, \hdelta\ and Mg II h\&k emission for the flare on the 8th May 2022, as traced by their amplitudes. As in Fig.\,\ref{fig:uv_flare} we have assumed the flare came from the primary star in the system. The line emission is for the combined system. Note the extended decay of the line emission relative to the \tess\ lightcurve. As in Fig.\,\ref{fig:uv_flare}, the optical and UV line emission here is for the combined system, resulting in reduced amplitudes. However, the morphology will be unchanged.}
    \label{fig:uv_flare_balmer_lines}
\end{figure}

\begin{figure}
    \centering
    \includegraphics[width=\columnwidth]{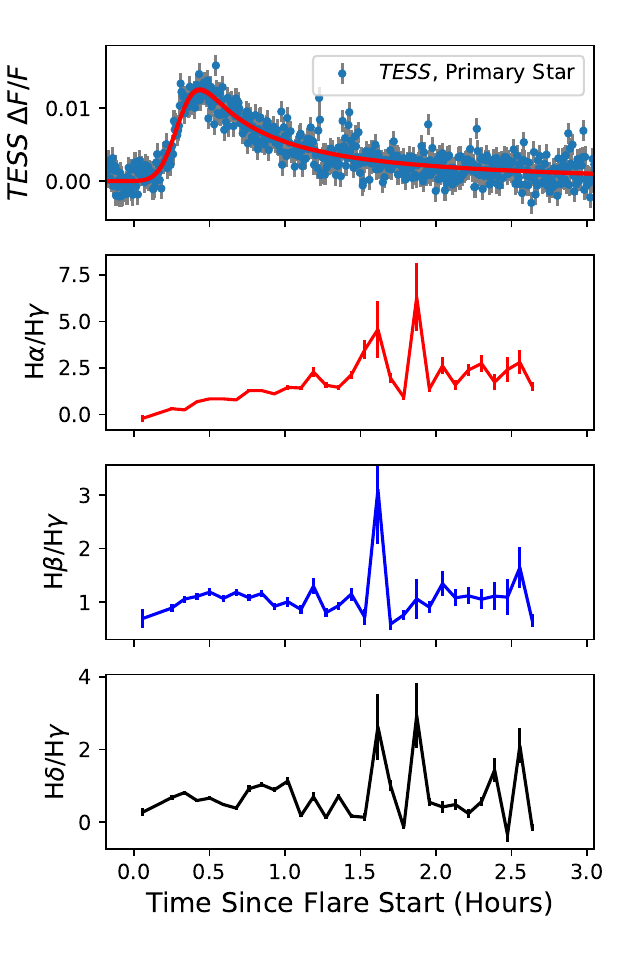}
    \caption{The Balmer decrements for \halpha, \hbeta\ and \hdelta\ for the flare on the 8th May 2022. As in Fig.\,\ref{fig:uv_flare} we have assumed the flare came from the primary star in the system. Note the increasing value of \halpha\ /\hgamma\ during the flare decay, due to the higher order emission fading faster than that from \halpha. We discuss this in Sect.\,\ref{sec:balmer_decrement}.}
    \label{fig:balmer_decrement}
\end{figure}

We used our observations to test the NUV predictions of flare models calibrated using \tess\ photometry. \citet{Jackman22} tested four flare models (and two sub-models) using \tess\ and archival \galex\ photometry. They found that the 9000\,K blackbody model underestimated the \galex\ NUV (1771-2831\AA) flare flux for partially-convective M stars by a factor of $2.7\pm0.6$. We calibrated the 9000\,K to the \tess\ amplitude measured during the \swift\ visits. We integrated both the 9000\,K blackbody and the \swift\ flare-only spectrum in the 2200--3000\AA\ range. We calculated that for our flare decay observations, the 9000\,K overestimated the NUV emission by a factor of \textcolor{black}{2.2} regardless of which star the flare came from. This value is at odds with previous works that have found this model to underestimate the total UV flux from flares \citep[e.g.][]{Jackman22,Brasseur23}. This suggests that the bulk of the NUV emission is emitted during the rise and peak phases of flares. We have investigated this for other flares in our sample in Sect.\,\ref{sec:results_models} and discuss the implications of this in Sect.\,\ref{sec:discuss_nuv_models}.

\subsubsection{Optical Spectra} \label{sec:optical_spectra}

We have used our simultaneous ground-based spectroscopy to study how the optical line emission changed throughout this event. We used our flux-calibrated spectra to construct lightcurves of the emission from the \halpha, \hbeta\ and \hgamma\ lines. These lightcurves are shown with the \tess, B band and Mg II h\&k emission in Fig.\,\ref{fig:uv_flare_balmer_lines}. We integrated over the \halpha, \hbeta\ and \hgamma\ lightcurves to calculate energies of \textcolor{black}{$1.6\pm0.1\times10^{32}$}, \textcolor{black}{$1.1\pm0.1\times10^{32}$} and \textcolor{black}{$9.2\pm0.1\times10^{31}$} ergs respectively. 

We can see in Fig.\,\ref{fig:uv_flare_balmer_lines} that both the Balmer and Mg II h\&k emission lines appear to have a longer decay than the continuum-dominated white-light emission. The \halpha\ emission in particular does not return to its pre-flare level before our observations finish. The long decay in \halpha\ relative to white-light photometry has been seen in previous flare studies \citep[e.g.][]{Kahler82, Kowalski13, Namizaki23}. It has been attributed to both a non-linear relation between the emergent continuum and broadened \halpha\ emission from lower chromospheric layers \citep[heated by the chromospheric condensation][]{Namekata20}, and the addition of red-shifted \halpha\ emission from post-flare loops. We can see that the \hbeta\ and \hgamma\ show different time evolution to the \halpha\ emission. This was observed by \citet{Kowalski13} in optical spectroscopy of flares from active M stars, and it was linked to higher order emission coming from higher density regions which cool faster than those associated with lower order emission. \textcolor{black}{A shorter response time in the \hbeta\ line relative to \halpha\ line has also been observed in Solar flares, where it has been linked to lower enhancement above the background level and differences in the line formation \citep[][]{Capparelli17, Pietrow23}.}

We also used our lightcurves to compare how the Mg II emission changes relative to the total and pseudo-continuum NUV flux. We found that the total flux drops by \textcolor{black}{49} per cent between the first and second \swift\ visit, while the pseudo-continuum and Mg II fluxes drop by 55 and 32 per cent respectively. 
This shows that the Mg II emission evolves on a longer timescale than the NUV continuum. 

Our measurement of the \hgamma\ emission also provides us with an opportunity to constrain the NUV continuum emission at the peak of the flare. \citet{Kowalski13} measured a linear relation between the ratio of flux in the \hgamma\ line and the continuum around 4170\AA\, (C4170) and the strength of the Balmer jump at 3615\AA. They found that this relation was present across flares of different morphologies. We used our optical spectra at the flare peak to measure the flare-only flux in the \hgamma\ line and the C4170 integration region defined by \citet{Kowalski13}. Using these with Eq.\,5 from \citet{Kowalski13} we estimated that the Balmer continuum increased the flux at 3615\AA\ by a factor of 2.3 the flare peak. Extrapolating the Balmer \textcolor{black}{continuum}  
linear relation measured during the first flare visit to 3615\AA\ gives a value of $2.3\pm0.1$. The strength of Balmer jump and continuum likely stayed constant between the flare peak and the decay phase sampled during the first \swift\ visit. This is likely due to our flare having a low impulsivity, making it comparable to the gradual flare events from \citet{Kowalski13} than more impulsive events with rapid rise phases. Two of the gradual flares observed by \citet{Kowalski13} exhibited similar Balmer jumps at their peak and decay, albeit with greater values ($>3$) than that measured here. Further simultaneous optical and NUV flare spectroscopy will help elaborate on this relation and determine the Balmer contribution at all flare phases. 

\textcolor{black}{We also calculated the ratio of the \halpha\ and \hbeta\ emission during our flares with optical spectra. We measured average (within single flares) \hbeta/\halpha\ ratios between \textcolor{black}{0.6} and \textcolor{black}{2.1} for the flares with complete coverage in Tab.\,\ref{tab:spectral_energies}. Reports of simultaneous detection of \halpha\ and \hbeta\ emission in Solar flares in the literature are rare, however the \hbeta/\halpha\ ratio has been investigated through simulations as a possible tracer of injected energy, particularly at different footpoints of the same flare event \citep[e.g.][]{Kasparova09,Capparelli17}.  \citet{Capparelli17} measured the \halpha\ and \hbeta\ emission associated with multiple footpoints of a C3.3 Solar flare. They isolated emission from three footpoints. At the two strongest footpoints (which showed the greatest level of emission) the ratio of the two emission lines stayed roughly constant between 2.0-2.5. At the weaker footpoint the ratio stayed around a value of 1.25, suggesting the lower level of energy deposition resulted in less \hbeta\ emission relative to \halpha. In the stellar context, \citet{Maehara21} measured values of the \hbeta/\halpha\ ratio between below 0.4 and a value of 1 for flares from the M4.5Ve star YZ CMi. They noted that according to \citet{Drake80} this ratio depends on the electron density, temperature and optical depth. However, they also suggested that the lower optical depth of the \hbeta\ line meant that flares with smaller \hbeta/\halpha\ ratios had lower flare region electron densities. Our measured ratios span a range similar to these values to the different footpoints studied by \citet{Capparelli17} and different flares studied by \citet{Maehara21}. Therefore, the range of our values may be due to changes in the energy deposition rate, or different electron densities in the flare regions, or most likely a combination of both.}

\subsection{Balmer Decrement} \label{sec:balmer_decrement}
\textcolor{black}{We used our spectra to calculate the \hbeta\ Balmer decrement for our flares. The Balmer decrement is the ratio of the background-subtracted flux in the Balmer lines to that of \hgamma\ \citep[e.g.][]{Hawley91}. \citet{Kowalski17} noted that M dwarf flares have typically measured values of \hbeta/\hgamma\ and \hdelta/\hgamma\ of 1.0--1.2 and 0.75--1.05 respectively. We calculated the average values for \hbeta/\hgamma\ for our flares and show these in Tab.\,\ref{tab:balmer_decrements}. We calculated \hdelta/\hgamma\ for the May 8th flare and present this analysis below.}  
\textcolor{black}{We measured \hbeta/\hgamma\ values between \textcolor{black}{0.9} and \textcolor{black}{1.2}. These values are broadly in line with measurements from previous M dwarf flare observations. Our average values do not vary strongly between flares in our sample, implying similar electron densities in the Balmer line forming regions between events \citep[e.g.][]{Garcia02}.}

\textcolor{black}{We measured how the \halpha/\hgamma, \hbeta/\hgamma\ and \hdelta/\hgamma\ ratios changed during the May 8th flare. The time evolution of these decrements are shown in Fig.\,\ref{fig:balmer_decrement}. We measured average values of 1.0 and 0.7 for \hbeta/\hgamma\ and \hdelta/\hgamma\ respectively. This value for \hdelta/\hgamma\ is lower than those previously measured for M dwarf flares, however we note that this is an average value while many previous reported only during the rise or decay phases \citep[e.g.][]{Allred06}. }
\textcolor{black}{Around the flare peak, our measured average value rises to 0.75, more in line with previous observations \citep[e.g.][]{Kowalski17}. We note that the \hbeta/\hgamma\ ratio appears to rise and decay with the change in the white-light then remain constant during the flare peak, suggesting the charge density of the emitting region is not varying strongly around the flare peak \citep[e.g.][]{Garcia02}. In contrast, we can see that the \halpha/\hgamma\ ratio increases throughout the flare. The observed increase in this ratio during the decay phase is due to the more rapid decrease in emission from higher order Balmer lines as the decay begins \citep[e.g.][]{Hilton10}. We note as well, that as discussed in Sect.\,\ref{sec:optical_spectra} the \halpha\ emission during the decay may have additional contributions from post-flare loops, whereas the higher order Balmer emission is expected to arise from lower chromospheric layers heated by condensations \citep[e.g.][]{Namekata20}. This difference in emitting regions will further increase the \halpha\ emission relative to \hgamma\ during the decay, as observed here. }

\subsection{\tess\ and Ground-based Spectra} \label{sec:results_optical}
We used our simultaneous optical spectroscopy to measure the Balmer line emission energies of flares also detected with \tess.  
During our analysis we masked any results with lower 1$\sigma$ uncertainties (16th percentile) greater than half of the final value. We did not find any evidence for flares that were only detectable through changes in line emission \citep[e.g. non-white light flares][]{Maehara21}. 

We measured energies in the \halpha\ line between \textcolor{black}{$2.7\pm0.5\times10^{30}$} and \textcolor{black}{$1.6\pm0.1\times10^{32}$} erg, \hbeta\ line between \textcolor{black}{$5.5\pm1.4\times10^{29}$} and \textcolor{black}{$1.1\pm0.1\times10^{32}$} and \hgamma\ between \textcolor{black}{$4.4\pm1.5\times10^{29}$} and \textcolor{black}{$9.2\pm0.1\times10^{31}$} ergs. As part of our analysis we measured the ratio between the energy from each emission line and the bolometric energy from a 9000\,K blackbody spectrum normalised to the flare amplitude in the \tess\ lightcurve. We did this twice, each time assuming each flare came from each star. For flares where we had complete spectroscopic coverage, we measured \halpha\ ratios between \textcolor{black}{1} and \textcolor{black}{4} per cent, \hbeta\ ratios between \textcolor{black}{1} and \textcolor{black}{5} per cent and \hgamma\ ratios between \textcolor{black}{1} and \textcolor{black}{3} per cent. The full list of flare energies and corresponding ratios are given in Tab.\,\ref{tab:spectral_energies}. 

\begin{table*}
    \centering
    \begin{tabular}{c|c|c|c|c|c|}
         Flare Start (TBJD) & Flare End (TBJD) & \halpha\ (erg) & \hbeta\ (erg) & \hgamma\ (erg) & \multicolumn{1}{|p{1cm}|}{\centering Complete\\ Coverage?} \\
         \hline
         2685.5216791737776 & 2685.58786429838 & N/A & $7.7\pm0.7\times10^{30}$ (0.01, 0.01) & N/A & Y \\
         2686.605896348823 & 2686.6368098146904 & N/A &  $7.1\pm0.6\times10^{30}$ (0.05, 0.05) & N/A & Y \\
         2686.7257868338374 & 2686.760732490905 & $3.6\pm1.1\times10^{30}$ (0.01, 0.01) & $7.4\pm0.7\times10^{30}$ (0.03, 0.02) & N/A & Y \\
         2689.9750280959197 & 2690.170167184171 & $1.3\pm0.6 \times 10^{32}$ (0.09, 0.09) & $8.0\pm0.1\times10^{31}$ (0.05, 0.05) & $6.7\pm0.1\times10^{31}$ (0.05, 0.04) & N \\
         2691.3765315668534 & 2691.405824710233 & $9.7\pm1.0\times10^{30}$ (0.01, 0.01) & $1.2\pm0.1\times10^{31}$ (0.01, 0.01) & $1.3\pm0.1\times10^{31}$ (0.02, 0.02) &  N \\
         2706.731994783021 & 2706.798031347259 & $2.7\pm0.1\times10^{31}$ (0.02, 0.02) & $2.1\pm0.1\times10^{31}$ (0.02, 0.02) & $2.0\pm0.1\times10^{31}$ (0.01, 0.01) & N \\
         2706.86020620054 & 2706.8787261972006 & $1.0\pm0.1\times10^{31}$ (0.02, 0.02) & $7.3\pm0.4\times10^{30}$ (0.01, 0.01) & $6.4\pm0.3\times10^{30}$ (0.01, 0.01) & N \\
         2708.3787735073875$\dagger$ & 2708.456550520614 & $1.6\pm0.1\times10^{32}$ (0.04, 0.04) & $1.1\pm0.1\times10^{32}$ (0.03, 0.03) & $9.2\pm0.1\times10^{31}$ (0.02, 0.02) & Y \\ 
         2708.7799270056357 & 2708.7919639265756  & $2.7\pm0.5\times10^{30}$ (0.01, 0.01) & $2.1\pm0.3\times10^{30}$ (0.01, 0.01) & $2.2\pm0.3\times10^{30}$ (0.01, 0.01) & Y \\
         2708.8368325096153 & 2708.8498997561073 & $3.9\pm0.5\times10^{30}$ (0.02, 0.02) & $3.1\pm0.4\times10^{30}$ (0.02, 0.02) & N/A & Y \\
         2724.6485486892016 & 2724.7761177149214 & $4.6\pm0.2\times10^{31}$ (0.03, 0.03) & $4.2\pm0.1\times10^{31}$ (0.03, 0.03) & $3.9\pm0.1\times10^{31}$ (0.03, 0.03) & Y \\
         2726.9616612145683 & 2726.9662936233335 & N/A & $5.5\pm1.4\times10^{29}$ (0.01,  0.01) & $4.4\pm1.3\times10^{29}$ (0.01, 0.01) & Y \\
         2742.9874324842285 & 2743.034828425528 & $5.2\pm0.7\times10^{30}$ (0.01, 0.01) & N/A & N/A & N 
    \end{tabular}
    \caption{Energies of flares detected with both \tess\ and ground-based spectra. The values in brackets are the ratio between the energy from the emission line and the bolometric energy from a 9000\,K blackbody normalised to the flare amplitude in the \tess\ bandpass, for the primary and secondary star respectively. We have only included flares where at least one energy calculation had a uncertainty below 50 per cent. N/A indicates where a spectral either did not have coverage or the calculated energy had an uncertainty above 50 per cent. The $\dagger$ symbol denotes the flare discussed in Sect.\,\ref{sec:result_may8}.}
    \label{tab:spectral_energies}
\end{table*}

\begin{table*}
    \centering
    \begin{tabular}{|c|c|c|c|c| }
         Flare Start (TBJD) & Flare End (TBJD) & \hbeta\ /\halpha & \hbeta\ /\hgamma &\multicolumn{1}{|p{1cm}|}{\centering Complete\\ Coverage?} \\
         \hline
         2685.5216791737776 & 2685.58786429838 & N/A & N/A & Y \\
         2686.605896348823 & 2686.6368098146904 & N/A & N/A & Y \\
         2686.7257868338374 & 2686.760732490905 & 2.1 & N/A & Y \\
         2689.9750280959197 & 2690.170167184171 & 0.6 & 1.2 &  N \\
         2691.3765315668534 & 2691.405824710233 & 1.2 & 0.9 & N \\
         2706.731994783021 & 2706.798031347259 & 0.8 & 1.1 & N \\
         2706.86020620054 & 2706.8787261972006 & 0.7 & 1.1 & N \\
         2708.3787735073875$\dagger$ & 2708.456550520614 & 0.7 & 1.2 & Y \\ 
         2708.7799270056357 & 2708.7919639265756 & 0.7 & 0.9 & Y \\
         2708.8368325096153 & 2708.8498997561073 & 0.8 & N/A & Y \\
         2724.6485486892016 & 2724.7761177149214 & 0.9 & 1.1 & Y \\
         2726.9616612145683 & 2726.9662936233335 & 2.0 & 1.2 & Y \\
         2742.9874324842285 & 2743.034828425528 & N/A & N/A & N 
    \end{tabular}
    \caption{\textcolor{black}{Average \hbeta\ /\halpha\ and \hbeta\ /\hgamma\ decrements of flares detected with both \tess\ and ground-based spectra.}}
    \label{tab:balmer_decrements}
\end{table*}

\subsection{Flare Temperatures} \label{sec:result_flare_teff}
We detected \textcolor{black}{seven} flares with both \tess\ and from the ground with B band photometry. We measured the temperatures of these flares following the methods discussed in Sect.\,\ref{sec:method_flare_teff} and their values are reported in Tab.\,\ref{tab:flare_teff}. For flares coming from the primary star, we measured average temperatures of between \textcolor{black}{6050 and 9480\,K}, average FWHM temperatures of between \textcolor{black}{6320 and 10710\,K}, and peak temperatures between \textcolor{black}{7100 and 17280\,K}. 
Our temperature measurements add to the growing gallery of studies showing that flares can exhibit a wide range of temperatures \citep[e.g.][]{Kowalski13,Howard20}, with all but one showing values below the typically assumed 9000\,K temperature \citep[e.g.][]{Maas22}.

We used our \tess\ observations to measure the impulse of each flare. We did this by dividing the amplitude in the \tess\ bandpass by the measured \thalf\ timescale. Previous studies have investigated whether the flare temperature is related to the flare impulse, namely if more impulsive flares have higher emission temperatures. We compared our sample to the \tess-EvryScope sample from \citet{Howard20}. We found that when expressed in terms of the impulse measured from the \tess\ lightcurves, our flares sit within the bulk of the \citet{Howard20} sample. 

\subsection{Testing Flare Models In The NUV} \label{sec:results_models}
In Sect.\,\ref{sec:result_may8} we tested the NUV prediction of the 9000\,K blackbody model for the flare observed on the 8th May. This model is used in flare studies for calculating bolometric energies of white-light flares. We found that it overestimated the NUV energy during the decay phase by a factor of \textcolor{black}{2.2}. Previous studies have noted that the 9000\,K blackbody model underestimates the UV emission of flares, notably at the peak \citep[e.g.][]{Kowalski19,MacGregor21}. 

To test this discrepancy further, we performed the same test for the other flares in our sample with simultaneous \tess\ and \swift\ observations. We do not have optical spectroscopy, or photometry other than from \tess, to constrain temperatures for these events. However, we were still able to use our observations to test the 9000\,K blackbody model. We took flares with simultaneous \tess\ and \swift\ observations and calculated the average NUV flare flux during the \swift\ visit. We did this using the lightcurves generated in Sect.\,\ref{sec:method_swift_energy}. We then measured the average optical flare amplitude at the same time using the \tess\ lightcurve. We then generated a 9000\,K blackbody curve and normalised it to give the required optical flux in the \tess\ bandpass. We then integrated the normalised 9000\,K blackbody curve in the NUV, using the same wavelength ranges as for our \swift\ lightcurves. The flux was then compared against the flare-only values measured from the \swift\ lightcurves.

\begin{figure}
    \centering
    \includegraphics[width=\columnwidth]{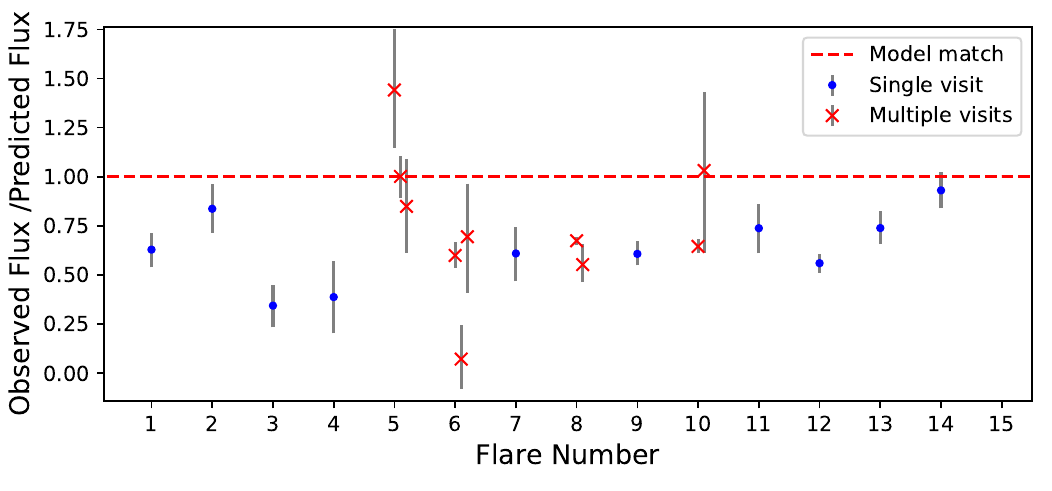}
    \caption{The ratio for the observed flare fluxes (2200--3000\AA) and those predicted by the 9000\,K blackbody model normalised using the flare amplitude in \tess\ photometry. These tests assumed the flare came from the primary star. The blue circles correspond to flares that were observed in a single \swift\ visit, while the red crosses are for flares observed in multiple consecutive visits. Red crosses for a given flare are offset for clarity. Note that the 9000\,K blackbody model overestimates the UV flux for all but one flare observation. We attribute this to these flares being observed during their decay phase. Flare number 8 is the 8th May flare discussed in Sect.\,\ref{sec:result_may8}.}
    \label{fig:model_test}
\end{figure}

The ratios of the observed and predicted NUV flare fluxes are shown in Fig.\,\ref{fig:model_test}. The calculated ratios ranged between 1.44 and 0.07 in the 2200-3000\AA\ lightcurves. However, 90 per cent of visits had calculated ratios below 1, showing that for the majority of observations the 9000\,K blackbody model overestimated the NUV flare flux. 

We can see in Fig.\,\ref{fig:model_test} that there are two flares (numbers 5 and 10) where the 9000\,K blackbody model underestimates the flux in a visit. The first of these, flare number 5, had a measured ratio of 1.44. Further investigation of flare 5 showed that the first visit occurred during the decay phase of a complex flare, but shortly before a second rise phase. We redid our tests using the pseudo-continuum lightcurve to see whether the result for this flare was driven by line emission. We found that the 9000\,K blackbody overestimated the pseudo-continuum flux for this visit, suggesting that the NUV emission during this visit was driven by the Mg II h\&k line. \citet{Kowalski19} used \hst\ NUV spectroscopy to study the temporal evolution of two flares from GJ 1243 and found in both cases that the continuum emission decayed faster than the line emission. This is similar to the behaviour previously seen for optical flare emission, something we discussed for the Balmer series in Sect.\,\ref{sec:result_may8}. It is likely our result in Fig.\,\ref{fig:model_test} is due to the Fe and Mg II h\&k line emission evolving slower than the continuum emission. Subsequent visits for the same flare occurred after a second reconnection event in the \tess\ lightcurve, and correspond to where the continuum emission is more dominant. The second event, flare number 10, had a ratio consistent with unity, telling us that in this scenario the 9000\,K accurately predicted the NUV flux. Analysis of this visit showed it occurred shortly after the flare peak, where the flare had begun cooling. Rerunning our test with the pseudo-continuum emission brought the measured ratio between the observed and predicted NUV flux below 1. Therefore our result in Fig.\,\ref{fig:model_test} was driven by the Mg II h\&k and Fe emission and that the flare temperature was well below 9000\,K.

\section{Discussion}
We have presented the results of a multi-wavelength observing campaign to study the optical and NUV emission of stellar flares from M stars. We used the \swift\ UV grism to obtain NUV spectra of flares and measure the NUV energy budget. Ground-based optical photometry and spectroscopy were combined with \tess\ photometry to measure the continuum temperatures of flares. 

\subsection{NUV energy budget}
We reported in Sect.\,\ref{sec:energy_result} that we measured 2200--3000\AA\ NUV flare energies between \textcolor{black}{$1.5\pm1.0\times10^{31}$} and \textcolor{black}{$1.2\pm0.2\times10^{33}$} ergs, and Mg II h\&k emission line energies between \textcolor{black}{$2.4\pm1.4\times10^{30}$} and \textcolor{black}{$1.8\pm0.4\times10^{32}$} ergs. We noted that these values are lower limits due to individual \swift\ visits only capturing parts of flares. 

We used the NUV spectra to measure the contribution of emission lines to the NUV energy budget. We found that the Mg II h\&k lines emitted between \textcolor{black}{9.3} and \textcolor{black}{24} per cent of the measured 2200-3000\AA\ NUV energy.  
\citet{Hawley07} measured that emission lines (both Mg II h\&k and Fe) could contribute up to 50 per cent of the total NUV (2300-3050\AA) flare flux, but were typically between 10 and 30 per cent. To better compare our data set to these previous results, we recalculated these ratios for the 2300--3000\AA\ wavelength range. We calculated that the Mg II h\&k line emitted between 10 and 24 per cent of the 2300--3000\AA\ NUV energy. To compare this with the values from \citet{Hawley07} we need to include the line emission from the Fe series as well. \citet{Hawley07} noted that the Mg II h\&k and Fe lines showed an almost equal division of energy during flares. In contrast, \citet{Kowalski19} measured Mg II emitting over double the Fe II flare energy during the peak of two flares. However, this difference between from \citet{Hawley07} may be due to a difference in the amount of lines used to measure the Fe II energy. \citet{Kowalski19} used \hst\ observations spanning 2510--2841\AA, missing Fe II emission lines in the 2300-2510\AA\ region. If we use the \citet{Hawley07} flux ratio to estimate the total contribution of lines to the flare energy budget, we estimate fractions between \textcolor{black}{20} and \textcolor{black}{48} per cent of the total 2300-3000\AA\ NUV energy, with 80 per cent of our observations having ratios below 40 per cent.  These results reinforce previous findings that the majority of the NUV energy emitted from flares is from the continuum emission, itself driven by the Balmer jump.

\subsection{Testing Flare Models} \label{sec:discuss_nuv_models}

In Sect.\,\ref{sec:result_may8} and \ref{sec:results_models} we tested the NUV predictions of the 9000\,K blackbody model for flares that had simultaneous \tess\ and \swift\ observations. We found that after using \tess\ photometry to calibrate the blackbody model, it overestimated the NUV emission of all but one of our flares. As mentioned in Sect.\,\ref{sec:result_may8} this result appears to be in conflict with previous studies that have found that the 9000\,K blackbody underestimates the NUV emission of flares \citep[e.g.][]{Kowalski19,Jackman22,Brasseur23}. \citet{Kowalski19} found that, when fit to the optical continuum, the 9000\,K blackbody model underestimated the NUV flux of flares observed with optical and \hst\ NUV spectroscopy by between factors of 2 and 3. This discrepancy was attributed to the contribution of the Balmer jump and NUV line emission, which increase the NUV emission above the optical continuum. \citet{Jackman22} used archival \galex\ data with contemporary \tess\ observations to test the UV predictions of flare models, and found the 9000\,K blackbody underestimated the \galex\ NUV emission of partially-convective M stars by a factor of $2.7\pm0.6$. \citet{Brasseur23} used contemporaneous \kepler\ and \galex\ observations to identify that the 9000\,K blackbody model underestimated the \galex\ NUV energy of all but the lowest energy flares in their sample. They found evidence that the discrepancy was dependent on the \galex\ flare energy, with higher energy flares exhibited greater fractions of their flux in the \galex\ NUV band. Similar behaviour was also identified by \citet{Jackman22}, who found that higher energy flares required greater UV energy correction factors to bring the predictions of optically-calibrated flare models in line with observations.
\citet{Brasseur23} noted that blackbodies with fixed flare temperatures of 18,000 and 36,000\,K could not explain the optical/UV ratios of the highest energy flares in their sample, highlighting the limitations of single-temperature blackbodies in describing multi-wavelength flare emission.

As stated above, we found the 9000\,K blackbody overestimated the NUV emission for 90 per cent of our simultaneous optical and NUV observations. This result is in contrast to the previously mentioned studies that found the 9000\,K blackbody model (and blackbodies of greater temperatures) underestimate NUV flare emission. We attribute this to our observations primarily being taken during the flare decay phase, when the emitting plasma cools. Studies testing flare models have either focused on the emission at the flare peak \citep[e.g.][]{Kowalski19}, or considered the total NUV energy budget from across all flare phases \cite[][]{Jackman22,Brasseur23}. It is believed that the bulk of flare UV emission occurs during the rise and peak phases of flares \citep[e.g.][]{MacGregor21}. It is in this region where the highest flare temperatures are achieved, with temperatures of 15,000 and up to 40,000\,K being measured in the optical and FUV \citep[e.g.][]{Loyd18,Froning19,Howard20}. In Sect.\,\ref{sec:method_flare_teff} we used two-colour \tess\ and B-band photometry to measure pseudo-continuum temperatures for flares from CR Dra. We measured peak flare temperatures between \textcolor{black}{7100} and \textcolor{black}{17280}\,K, FWHM temperatures between \textcolor{black}{6320 and 10710\,K} and global average temperatures between \textcolor{black}{6050 and 9480\,K}. These results add to the growing body of evidence that not only do flares exhibit their highest temperatures during the rise and peak, but the average temperatures of flares span a range of values, with many exhibiting temperatures below the typically assumed 9000\,K value \citep[e.g.][]{Maas22}. 

In order to reconcile our results with those of previous studies, we revisited the \tess\ lightcurves for the flares with simultaneous \tess\ and \swift\ observations. We used these lightcurves to estimate the average temperatures these flares needed to have to give NUV fluxes 2 to 3 times that predicted by a 9000\,K blackbody. We did this by calculating the fraction of the flare our \swift\ visits covered, and by how much the 9000\,K blackbody model overestimated the NUV flux during this time. We found that in order to have a NUV flux 2--3 times that of a 9000\,K blackbody, our flares required an average temperature of 13,000\,K outside of the \swift\ visits. However, when we assume a contribution of a factor of 3 in flux from the Balmer jump \citep[e.g.][]{Kowalski13}, the required flare temperatures decrease to the 9000-9500\,K temperature range. 

This finding is not a significant increase above the typically assumed 9000\,K value for the flare temperature. However, it should be noted that this value represents the average value outside of our \swift\ observations. 
It is likely that these flares will have had rise and peak temperatures above this value when we consider that the 9000\,K blackbody overestimating the NUV flux in our \swift\ observations highlights that the flare temperatures during the \swift\ visits must have been well below 9000\,K. 
We estimated the flare temperatures required to give the same NUV emission as a 9000\,K blackbody and these are given in \textcolor{black}{Tab.\,\ref{tab:temp_values}}. The temperatures listed in \textcolor{black}{Tab.\,\ref{tab:temp_values}} indicate upper limits for the flare temperatures of events we have detected. 
If we assume the Balmer continuum and UV lines contribute the same amount of flux as the optical continuum, then this means all of our detected flares had temperatures below 7586\,K during our \swift\ observations. These values are consistent with the lower end of our measured average flare temperatures in Tab.\,\ref{tab:flare_teff}. 
\citet{Kowalski13} measured Balmer \textcolor{black}{jump ratios} of up to 3 and even 4 during the decay phase of flares, showing the Balmer jump tripled or quadrupled the NUV flux. This suggests that 
our flares may have continuum temperatures below 6565 or 6953\,K, as emission lines contributed flux equal to twice or three times the extrapolated optical continuum. These results imply that flares from CR Dra regularly exhibit temperatures below 9000\,K during their decay phases. This will have the effect of increasing the temperature further during the flare rise. Our results not only show the limitations of single-temperature blackbody models for predicting UV emission, but also highlight how the accuracy of flare models also depends on the flare phase.

\subsubsection{The Balmer Series}
In Sect.\,\ref{sec:results_optical} we calculated the energies emitted by the \halpha, \hbeta\ and \hgamma\ emission lines. The measured energies are given in Tab.\,\ref{tab:spectral_energies}, along with the ratio of these energies to the bolometric energy of a 9000\,K blackbody. We can see for the flares that were observed in their entirety that these emission lines provide flux up to 9 per cent of the bolometric energy predicted by a 9000\,K blackbody normalised to the flare amplitude in the \tess\ bandpass. \citet{Maehara21} measured that the \halpha\ and \hbeta\ contributed fluxes equivalent to between 5 and 20 per cent of the continuum emission in the \tess\ bandpass itself. \citet{Hawley91} measured that emission lines across optical and ultraviolet wavelengths contributed 9 per cent of the total optical and UV flux of the 1985 Great Flare from AD Leo, and Hydrogen lines (including higher order than \hgamma) specifically contributed approximately 5 per cent of the total flux. \citet{Kowalski13} later measured ratios between the combined \halpha, \hbeta\ and \hgamma\ emission and the total optical flux at the flare peak up to 6 per cent for active M stars. These results show that current models miss flux at optical wavelengths that is important for precise calculations of bolometric flare energies. 

At wavelengths shorter than \hgamma\ the Balmer series has even greater importance, something highlighted by our result for the flare from the 8th May in Sect.\,\ref{sec:result_may8}. 
We used the simultaneous \tess\ and B band photometry to measure flare temperatures of $5860^{+80}_{-90}$\,K \textcolor{black}{($5880^{+80}_{-90}$)}\,K and $5540^{+230}_{-220}$\,K \textcolor{black}{($5580^{+230}_{-220}$)}\,K during the flare decay. 
Studies using the two-colour temperature to predict the UV emission would estimate a 2200-3000\AA\ flux up to \textcolor{black}{0.1} times that from a 9000\,K blackbody. However, as shown in Fig.\,\ref{fig:uv_flare}, this measured flare temperature underestimated the 2200-3000\AA\, flare flux by factors of \textcolor{black}{$4.2\pm0.2$} ($4.1\pm0.2$) and \textcolor{black}{$4.1\pm0.9$} ($4.1\pm0.9$) respectively.
We attributed this to the flux contribution from the Balmer jump and the Fe and Mg II h\&k emission lines. We measured a Balmer \textcolor{black}{continuum ratio} of \textcolor{black}{$2.6\pm0.1$} (\textcolor{black}{$2.6\pm0.1$}) at 3000\AA\ during the first \swift\ visit for the 8th May flare, which decreased to $1.8\pm0.3$ ($1.7\pm0.3$) during the second visit. 

\citet{Kowalski13, Kowalski19} measured the Balmer \textcolor{black}{jump ratio} for flares from active M stars and found that the size of the Balmer jump (relative to the continuum) increased during the decay phase of flares, as the emitting area cooled. They also found that less impulsive flares had cooler optical temperatures and larger Balmer jumps at their peaks. Recent studies have used RHD modelling to reproduce rise, peak and impulsive decay phase spectra of short timescale impulsive flare events, such as the 1985 Great Flare from AD Leo \citep[][]{Hawley91}. 
These models require high energy electron jets that heat deep chromospheric layers to reproduce flare colour temperatures of 9000-11000\,K and Balmer jump ratios of $\approx$ 2 \citep[][]{Kowalski22,Kowalski23}. These electron jets utilise higher energy cut offs than those typically seen in Solar flares to overcome thermalisation of electrons in the upper chromosphere and allow these jets to adequately heat deeper chromospheric layers \citep[e.g.][]{Kowalski17}. \citet{Kowalski22} used a two-component RHD model to simulate the Hydrogen recombination radiation of the Great Flare from AD Leo, with the lower beam flux ($\sim10^{11} \mathrm{erg\,cm^{-2}\,s^{-1}}$) having a filling factor/ emitting area 10 times the higher beam flux ($\sim 10^{13} \mathrm{erg\,cm^{-2}\,s^{-1}}$). The two RHD model components were likened to a bright flare kernel(s) and extended ribbons in the flaring atmosphere \citep[e.g.][]{Heinzel14}. 
\citet{Kowalski22} also noted that two-component RHD models can also reproduce flare spectra with larger Balmer jumps such as the one seen in \citet{Kowalski19} and in this work. However, they noted the lower beam flux required a filling factor 90 times the higher energy beam, implying the flare emission during the decay is driven by extended areas such as flare ribbons.

Our results add further proof that the Balmer jump must be considered when modelling the UV emission of flares. If all flare decays exhibit a similar Balmer \textcolor{black}{jump ratio}, then the NUV emission will be greater than previous studies predicted despite the lower flare temperatures.

\begin{table}
    \centering
    \begin{tabular}{|c|c|}
         Line Factor & Flare Temperature (K) \\
         \hline
         0 & 9000 \\
         1 & 7586 \\
         2 & 6953 \\
         3 & 6565 \\
         4 & 6290 \\
    \end{tabular}
    \caption{Blackbody temperatures required for a flare with different line factors to give the same 2200--3000\AA\ emission as a 9000\,K blackbody. All blackbodies were fixed to give the same flux in the \tess\ bandpass. Here the line factor is the amount of emission from the Balmer jump and UV lines relative to the extrapolated optical continuum. A value of 1 means that UV lines and the Balmer jump provide the same amount of flux as the extrapolated optical continuum. As the 9000\,K blackbody model overestimated the NUV flux during \swift\ observations, these values provide upper limits for the temperatures of our detected flares, depending on the assumed contribution from line emission.}
    \label{tab:temp_values}
\end{table}

\section{Conclusions}
We conducted a multi-wavelength observing campaign to study flares from the active M binary CR Draconis. We did this using \tess, \swift\ and ground-based optical photometry and spectroscopy from a Professional-Amateur collaboration. 

We presented a flare detected with \tess, \swift\ and ground-based B-band photometry and optical spectroscopy. We used the simultaneous optical photometry and spectroscopy to calculate the temperature of the flare, measuring a peak value of $7100^{+150}_{-130}$\,K ($7130^{+150}_{-130}$\,K). We used our temperature model to calculate the predicted NUV emission during the \swift\ observations, and found it underestimated the 2200--3000\AA\ NUV emission by up to a factor of $4.2\pm0.2$ and the pseudo-continuum emission by up to a factor of $3.1\pm0.2$. We found that the pseudo-continuum emission decayed faster than the total NUV emission, which we identified was due to the Mg II h\&k emission evolving on a longer timescale. We also used our NUV grism spectroscopy with the \tess\ photometry to calculate the contribution of the Balmer jump during our flares. The flux from the Balmer jump increased the NUV continuum in our observations by up to a factor of $2.6\pm0.1$ at 3000\AA. \textcolor{black}{We measured the Balmer decrement of this flare, finding that the \halpha\ decrement increased during the flare, while the \hbeta\ and \hgamma\ decrement showed morphology in line with the \tess\ photometry.}

We used the simultaneous \tess\ optical and \swift\ near-ultraviolet observations to test the predictions of the commonly used 9000\,K blackbody flare model. We found that in the majority of our observations the model overestimated the total NUV emission of flares. This is in contrast to recent results that have found that this model underestimates the NUV flare flux. We attributed this to our observations primarily occurring during the flare decay, during which flare emitting regions cool. In one \swift\ visit the 9000\,K blackbody underestimated the NUV emission. However, after using our grism spectra to isolate the pseudo-continuum NUV emission we identified that the total flux in this case was likely driven by an extended phase of Mg II h\&k emission during the flare decay. 

Our observations show the capability of the \swift\ grism in studying the NUV emission of stellar flares and its use in testing flare models. Our results also highlight the role Pro-am collaborations have in improving our understanding the multi-wavelength characteristics of stellar flares. 

\section*{Acknowledgements}
This research has made use of the SVO Filter Profile Service (http://svo2.cab.inta-csic.es/theory/fps/) supported from the Spanish MINECO through grant AYA2017-84089. 
JAGJ acknowledges support from grant HST-AR-16617.001-A from the Space Telescope Science Institute, which is operated by the Association of Universities for Research in Astronomy, Inc., under NASA contract NAS 5-26555. 

This paper includes data collected by the \tess\ and \swift\ missions, which are publicly available from the Mikulski Archive for Space Telescopes (MAST). Funding for the \tess\ mission is provided by NASA's Science Mission directorate. This research was supported by NASA under grant number \textcolor{black}{80NSSC22K0125} from the \tess\ Cycle 4 Guest Investigator Program.

This research is based on observations made with the \textit{Neil Gehrels Swift Observatory}, obtained from the MAST data archive at the Space Telescope Science Institute, which is operated by the Association of Universities for Research in Astronomy, Inc., under NASA contract NAS 5–26555. This research was supported by NASA under grant numbers 80NSSC22K1493 and 80NSSC23K0311. We acknowledge the use of public data from the \swift\ data archive. 
We acknowledge with thanks the variable star observations from the AAVSO International Database contributed by observers worldwide and used in this research.

\section*{Data Availability}
All \tess, \swift\ and ground-based optically photometry data is publicly available. \tess\ photometry and target pixel files can be obtained from MAST. The \swift\ observations can be downloaded from the NASA archive. Ground-based photometry can be downloaded from the AAVSO International Database. Ground-based spectroscopy is available upon request to the observers.



\bibliographystyle{mnras}
\bibliography{references} 





\bsp	
\label{lastpage}
\end{document}